\documentclass[a4paper,12pt]{article}
\usepackage[top=3cm, bottom=3cm, left=3cm, right=3cm]{geometry} 
\usepackage{color}
\usepackage{rotating}
\usepackage{graphicx}
\usepackage{amsmath}
\usepackage{amssymb}
\usepackage{float}
\usepackage{subfig}
\usepackage{subfloat}
\usepackage{hyperref}
\def\displayandname#1{\rlap{$\displaystyle\csname #1\endcsname$}%
                      \qquad \texttt{\char92 #1}}

\begin{document}
\title{\bf{Characterization and Comparison of Glass Electrodes}}

\author{R. Kanishka\thanks{email:kanishka.rawat@saha.ac.in}~ \\
  V. Bhatnagar$^\dagger$ \\
  {\it $^*$APND, Saha Institute of Nuclear Physics, Kolkata 700064, India}\\
  {\it $^\dagger$Physics Department, Panjab University, Sector 14, Chandigarh 160 014, India} \\
}

\maketitle
%\emailAdd{kanishka.rawat@saha.ac.in}

\begin{abstract}{This paper presents the study on the characterization of glass electrodes, which are one of the main components of detectors like Resistive Plate Chambers (RPCs). The RPCs are being used in various ongoing High Energy Physics experiments, e.g., BELLE at KEK, CMS at LHC, and would be used in the near future experiments e.g., INO-ICAL in India. The characterization of glass electrodes has been done to understand the factors like quality of glass that can help in improving the detector's performance. The glass samples chosen were procured locally Asahi (A), Saint Gobain (S), Modi (M) that are easily available in Indian industry. The characterization includes the tests to study the optical, surface, physical, electrical properties, the composition of glass samples, and leakage currents. This paper adds new information to the existing body of research on the subject. Based on the techniques discussed in the paper a comparison of the measurements among the three different types of glass electrodes has been done. This study helps us to determine the best quality of glass that can be chosen for better operation of the detectors.}
\end{abstract}

%\keywords{Glass, Asahi (A), Saint Gobain (S), Modi (M), resistive, characterization, electrode, surface.}

\newpage

%\maketitle
%\flushbottom

\section{Introduction}
\label{rpcs_paper}

\begin{figure}[htbp]
\centering
\includegraphics[width=0.6\textwidth]{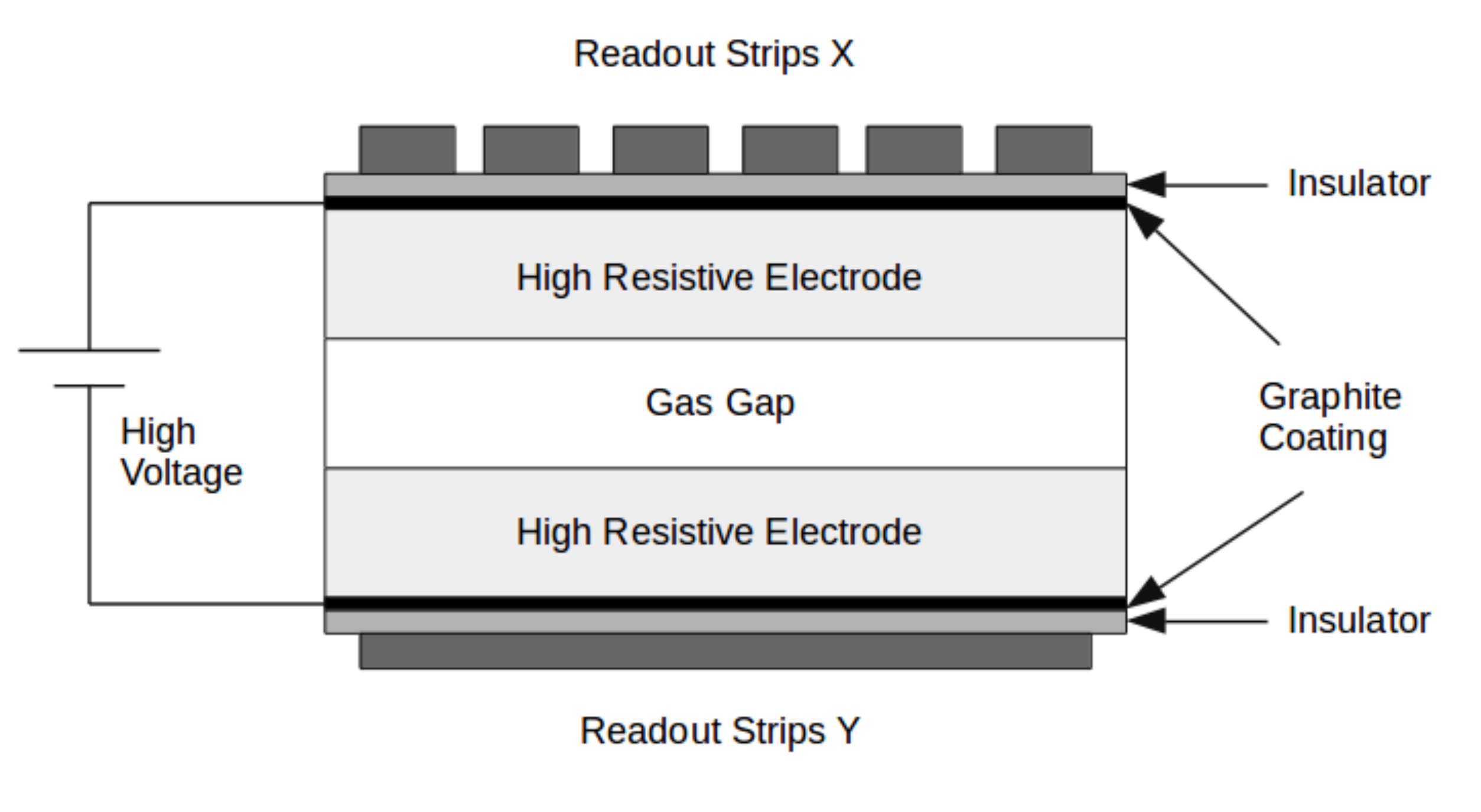}
\caption{A schematic of RPC detector.}
\label{rpc_view}
\end{figure}

RPCs \cite{cardarelli, Santonico, satya} are well established particle detectors used in various high energy experiments like BELLE \cite{belle}, CMS \cite{cms}, ATLAS and ALICE, and would also be used in the near future experiments like ICAL at INO \cite{ino}. The RPCs are low cost parallel plate gas detectors built using resistive electrodes made of bakelite or glass. They have excellent position and time resolutions \cite{cardarelli, fonte} that have been utilized for charged particle track detection. A standard procedure for the fabrication of an RPC is adopted by assembling two bakelite or glass plates having very high bulk resistivity, and has been discussed in section~\ref{assemblyP}. Electrical signals get induced on read-out boards (pick-up panels). These boards are oriented in orthogonal directions and placed at the outer surfaces of the electrodes. The pick-up panels are separated from the graphite coated glass RPC with the help of an insulating material (mylar). Figure~\ref{rpc_view} shows the schematic of an RPC detector.

Depending on the operating voltage and the gas mixture used, the RPC can be operated in an {\bf Avalanche} or a {\bf Streamer Mode} \cite{carda1}. In the avalanche mode, an incoming ionizing particle (cosmic muon \cite{muon1}) produces primary charge in the gas mixture and further multiplication of the electrons takes place. In streamer mode, an increase in the gain of gas causes the photons to contribute so that the amount of ionization exceeds the Raether limit \cite{Gasik}. To ensure that RPCs work efficiently for many years, it is important to understand the components of the RPC. The bulk resistivity and surface smoothness of the glass material helps to understand the discharge formation that deteriorates the long term stability of the detectors. The rough surface of the electrodes can cause the electric field to vary significantly. Thus the surface smoothness plays a major role in choosing the type of electrode for the detector. It is an important factor in reducing spontaneous discharge that may cause irreversible damages to the detector, such as enhanced leakage currents. A discharge may also cause a permanent short-circuit between the two sides of the electrodes, and many such short-circuits may cause the RPCs to become non-operational. In high rate experiments ageing resistance, radiation hardness, and stability against discharges are the main criteria for the long-term operation of the detectors \cite{Gasik}. Therefore, to improve the stability of the RPCs, a characterization of the electrode material is crucial. This paper is organized as follows: section~\ref{char_glass} discusses the characterization of glass electrodes. Section~\ref{fabri} details the fabrication and characterization of RPCs. Finally, the conclusions are presented in section~\ref{diss_con}.

\section{Characterization of Glass Electrodes}
\label{char_glass}

The glass electrodes are one of the crucial components of RPC. A proper choice of electrode is required to help maintain the stability of the RPC operation. Therefore, detailed characterization studies were done for various float glass samples procured from different manufacturers from the domestic market which were available at a reasonable cost \cite{Meghna}. The following properties were studied \cite{kanish}:

\noindent
{\bf {Physical Properties}}: The mass was measured with a fractional balance. The volume and density of all the glass samples were measured. We observed no significant difference in the density of the glass samples \cite{sgobain}. The results are given in table~\ref{table1}.

\begin{table}[ht]

\centering 
\begin{tabular}{|c c c c|} 
\hline 
Samples & Volume(${cm^{3}}$) & Weight(g) & Density(${g/cm^{3}}$)\\ 
\hline                  
A & 1.85 & 4.53 & 2.45 \\
S & 2.78 & 6.89 & 2.48  \\
M & 4.35 & 10.63 & 2.44 \\
\hline 
\end{tabular}
\caption{The density measurements of A, S and M glass samples.} 
\label{table1} 
\end{table}

\begin{figure}[htbp]
\centering
   \includegraphics[height=.3\textheight]{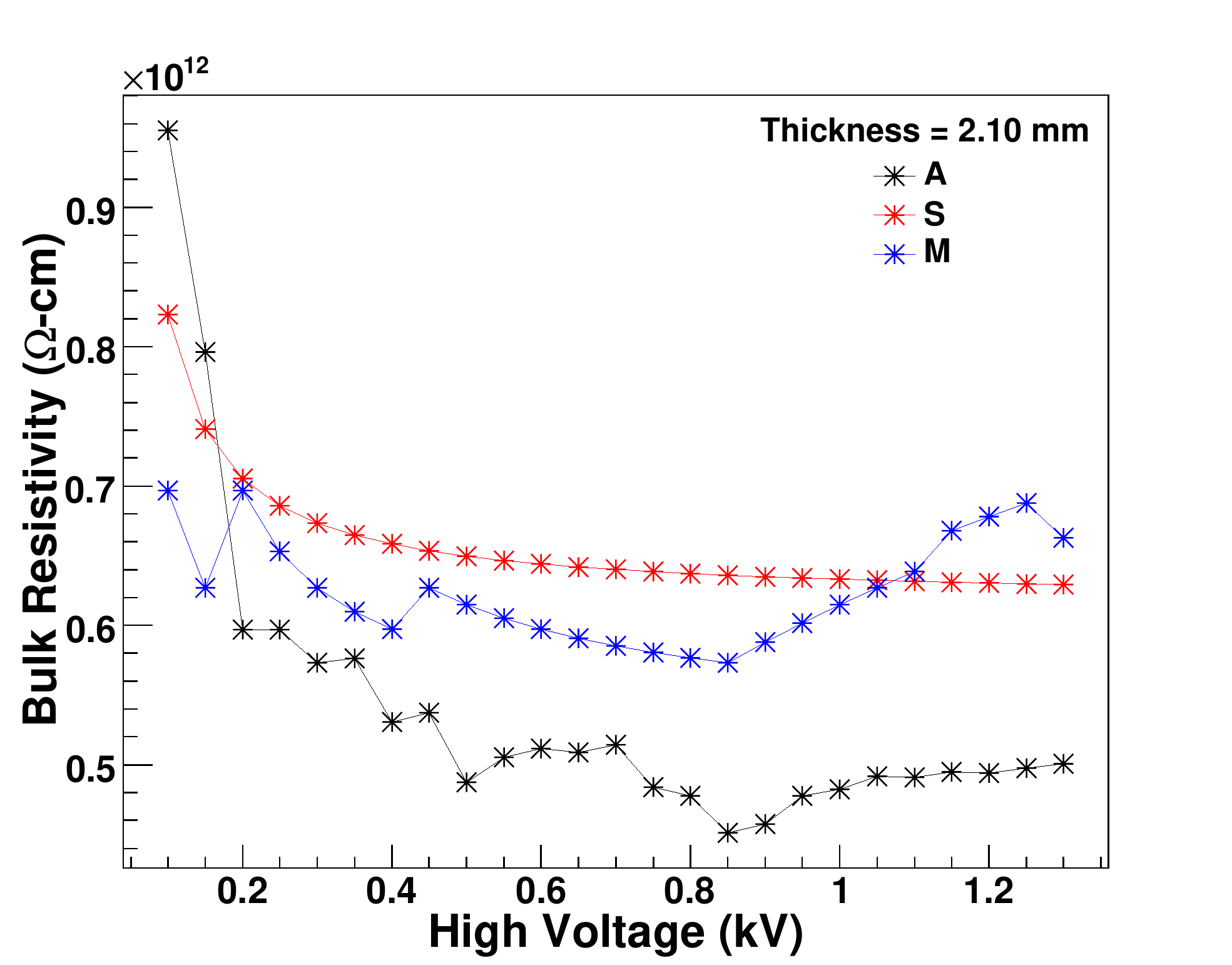}
  \caption{The bulk resistivity measurements of A, S and M glass samples.}
\label{fig:bulk1}
\end{figure}

\noindent
{\bf {Electrical Properties}}: The bulk resistivity measurements were done to determine the electrical properties of the electrodes. A non-smooth surface of the electrodes affects the detector performance by contributing towards the count rate and increase in the dark current, thereby affecting the efficiency of the RPCs. The bulk resistivity of the glass samples was measured using the two-probe method. It is one of the standard methods for the measurement of resistivity of samples like sheets and films of polymers with high resistivity. First, a small piece of aluminium foil was placed below the spring, and the glass sample was placed below the other probe. The leads were connected to high voltage power supply and a digital picoammeter. Then with the gradual increase of voltage, corresponding current values were noted. The measurements of the dimensions (length, breadth and thickness) of the samples were done to calculate the resistivity. Figure~\ref{fig:bulk1} shows the bulk resistivity of A, S and M glass samples (for samples having a thickness of 2.10 mm). The bulk resistivity of the samples was found of the order of $10^{11}$ $\Omega$-cm \cite{satya}. An advantage of high bulk resistivity of the electrode is that it helps to localize the excess charge and hence improves the performance of RPCs.

\begin{figure}[htbp]
\centering
 \includegraphics[height=.3\textheight]{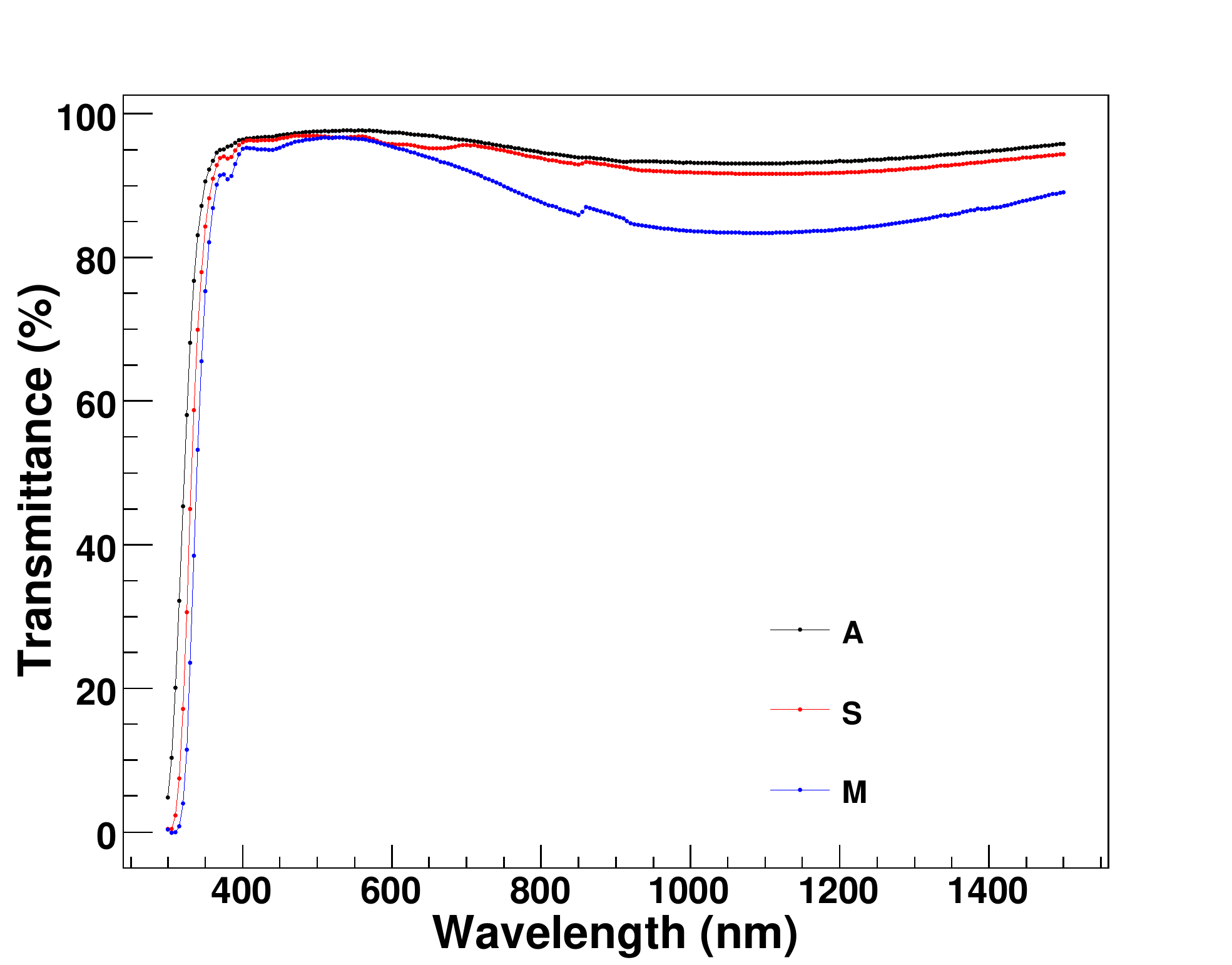}  
    \caption{UV-VIS transmittance test of A, S and M glass samples to check the quality of glass.}
\label{fig:bulk}
\end{figure}

\noindent
    {\bf {Optical Properties}}: The transmittance of various glass samples in the ultraviolet (UV) to visible light spectrum was tested. This test was done using Perkin Elmer's LAMBDA 750 UV/VIS/NIR and UV/VIS spectrophotometers \cite{cil} that have a resolution of 0.17-5.00 nm in the UV and visible region and 0.20-20.00 nm in the near infrared region. This spectrophotometer uses highly sensitive photomultiplier and Peltier cooled PbS detectors to provide a full range of UV/VIS/NIR region from 190 to 3300 nm. When radiation passes through a sample, some of it is absorbed by the molecules in the sample. Then a transition to an excited state takes place. When the molecules in the sample return to the ground state emission takes place in the form of radiation, heat, fluorescence, or phosphorescence. Therefore, when light is passed through a sample the intensity of incoming light, $I$, reduces to $I_{0}$ because of the loss of energy. The transmittance of a sample is defined as $I_{0}$/$I$. This test was performed to determine the general quality of glass, which is shown in figure~\ref{fig:bulk}. The transmittance is also used to check the surface smoothness. This is because when light passes through an uneven surface of glass a portion of light gets scattered too. Therefore the transmittance is further reduced. Figure~\ref{fig:bulk} shows the optical transmittance for all the glass samples. The A and S glass samples show better UV-VIS transmittance than M sample. This further motivated us to scan the surface of these samples to check the surface smoothness.

\begin{figure}[htbp]
\centering
  \includegraphics[height=.16\textheight]{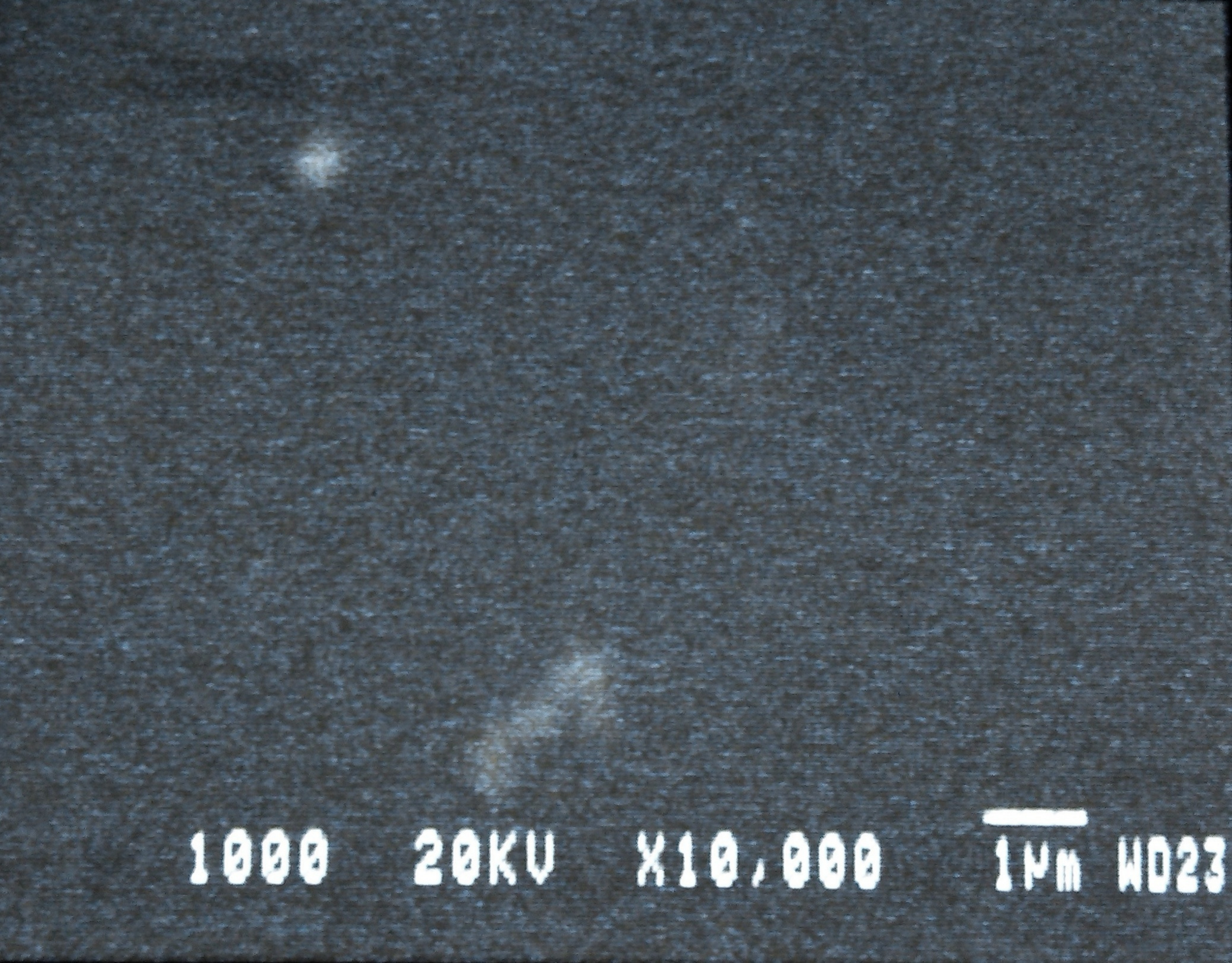}
\includegraphics[height=.16\textheight]{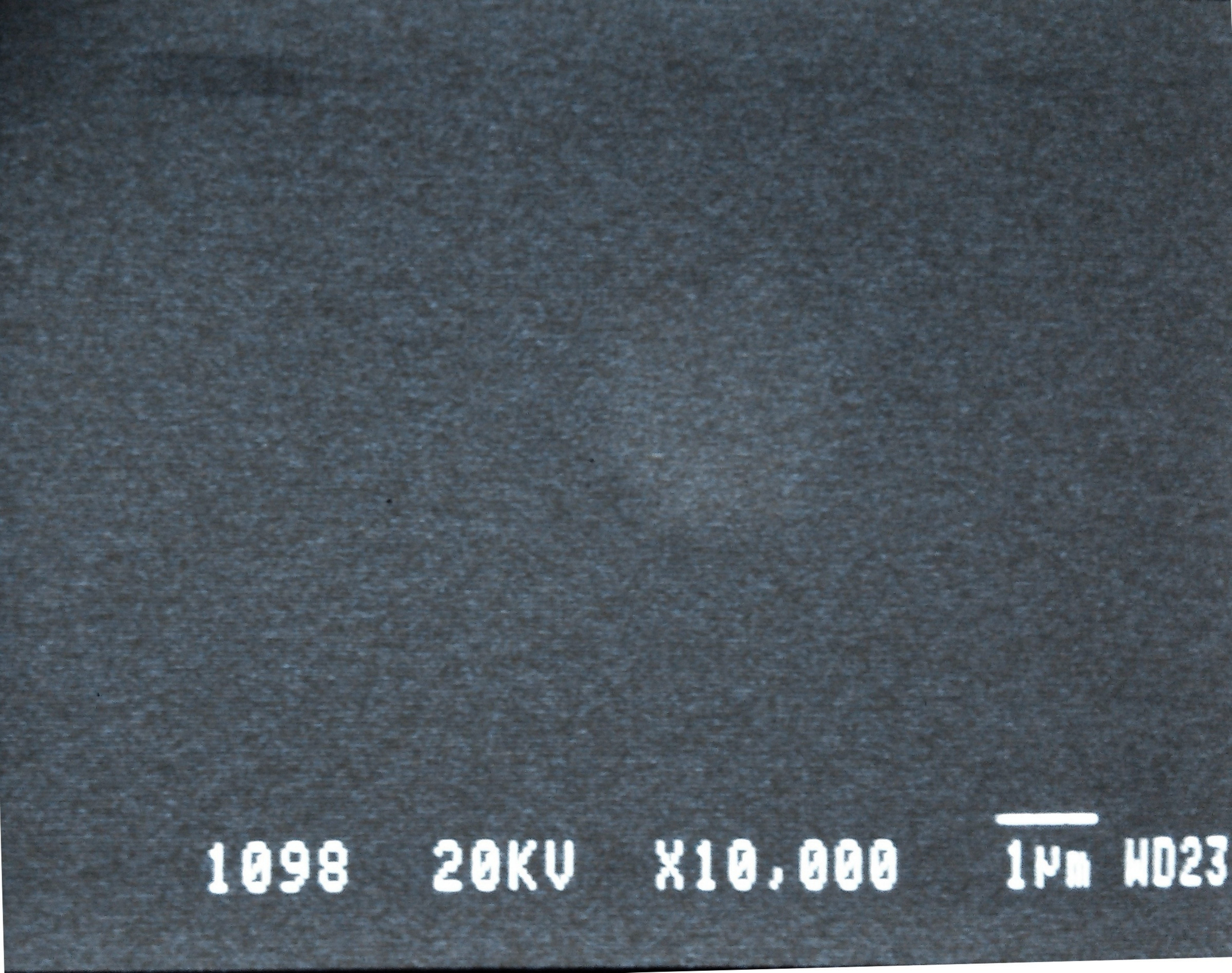}
\includegraphics[height=.16\textheight]{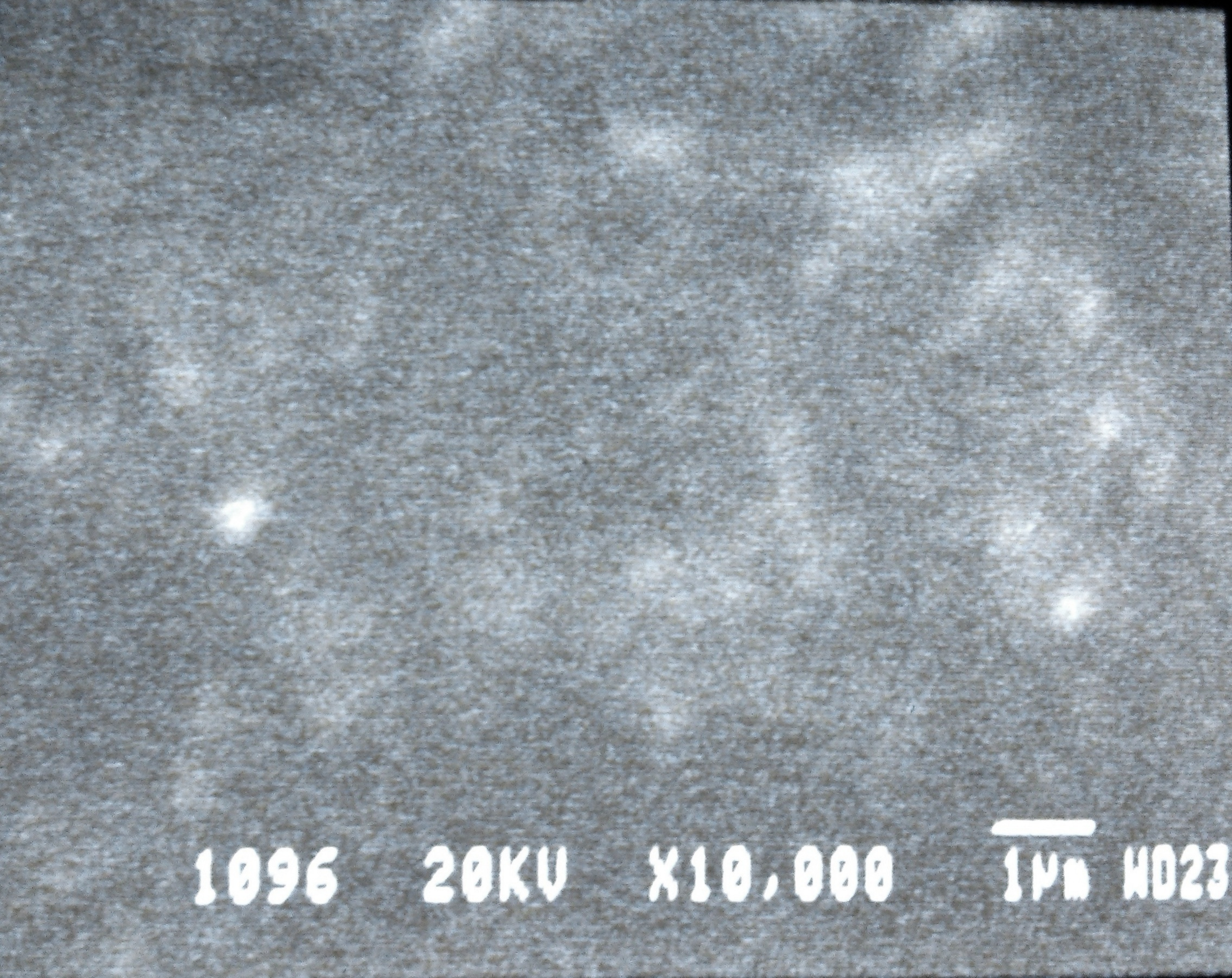}
  \caption{SEM scans of A, S and M glass samples respectively.}
\label{fig:SEM}
\end{figure}

\noindent
    {\bf {Surface Properties}}: The rough surface of the glass electrodes of RPC causes field emission which further affects the count rates and dark current. The surface quality of the electrode is crucial in reducing spontaneous discharges which affects the performance of the RPCs. Scanning Electron Microscopy (SEM) has been used to study and compare the surface quality of various glass samples. The SEM uses a beam of high-energy electrons to produce a variety of signals at the surface of the specimens due to the electrical coating on them. These signals then reveal the information about the sample like external morphology (texture), crystalline structure, chemical composition, and orientation of materials in the sample to study further. The SEM was done with digital Scanning Electron Microscope - JSM 6100 (JEOL) \cite{cil}. The glass samples were coated with a conductive coating of gold using a high-vacuum evaporation technique \cite{vaccum}. Figure~\ref{fig:SEM} shows the SEM of all the glass samples. For about 20 kV accelerating voltage of electrons and magnification of about 10000, we noticed that pores in case of A and S are fewer in number whereas a much larger number of pores were observed in the M glass sample. Thus A and S glass samples show a better quality of the surface than the M glass sample. The effect of an uneven surface of the M glass sample was further studied by characterizing the whole RPC. This will be discussed in section~\ref{viChar}.

\begin{figure}[htbp!]
\centering
 \includegraphics[height=12cm, width=14cm, scale=5]{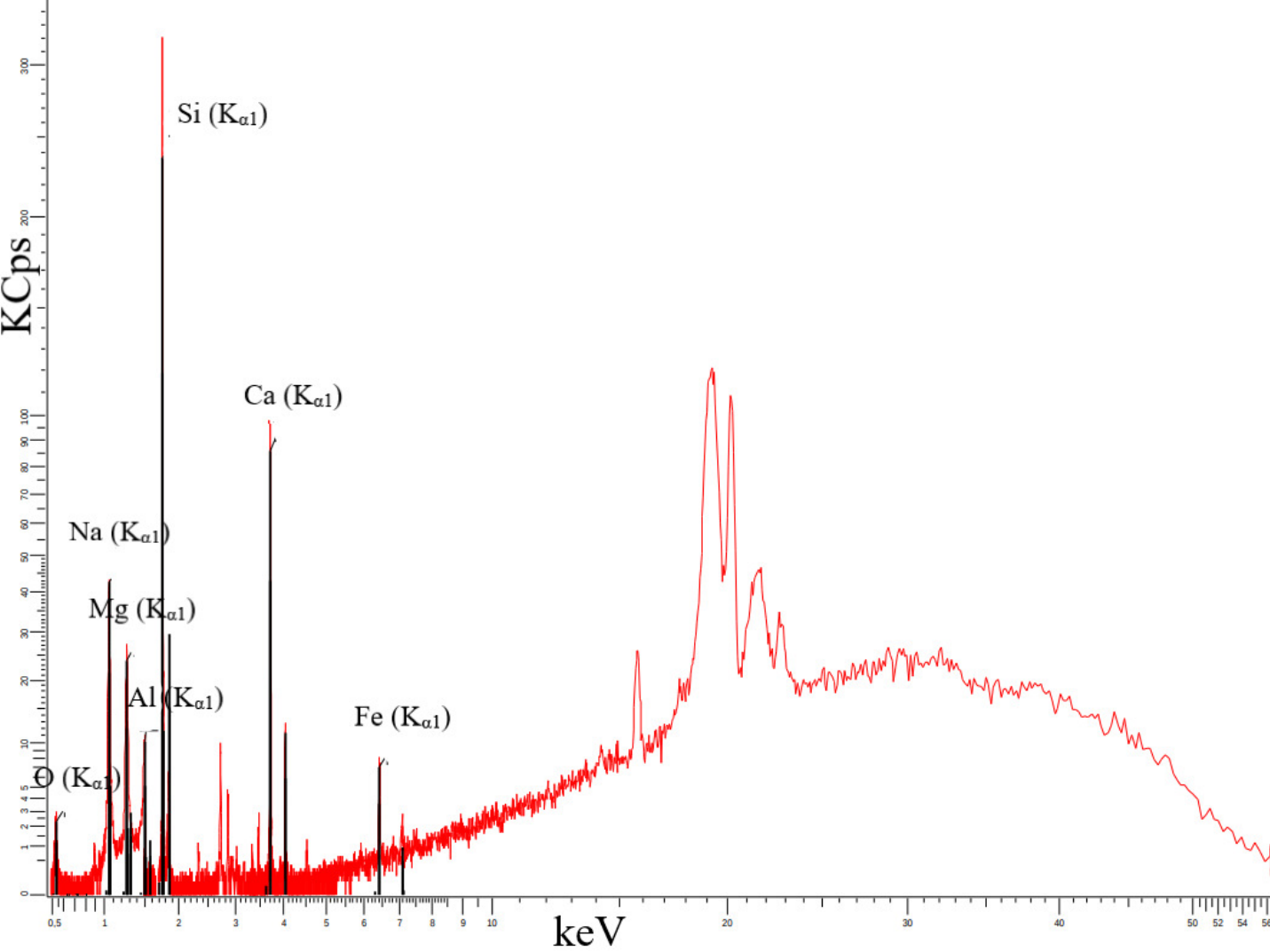}

  \caption{WD-XRF analysis of A, S and M glass samples respectively.}
\label{fig:WDXRF11}
\end{figure}

\begin{figure}[htbp]
\centering
\includegraphics[height=.3\textheight]{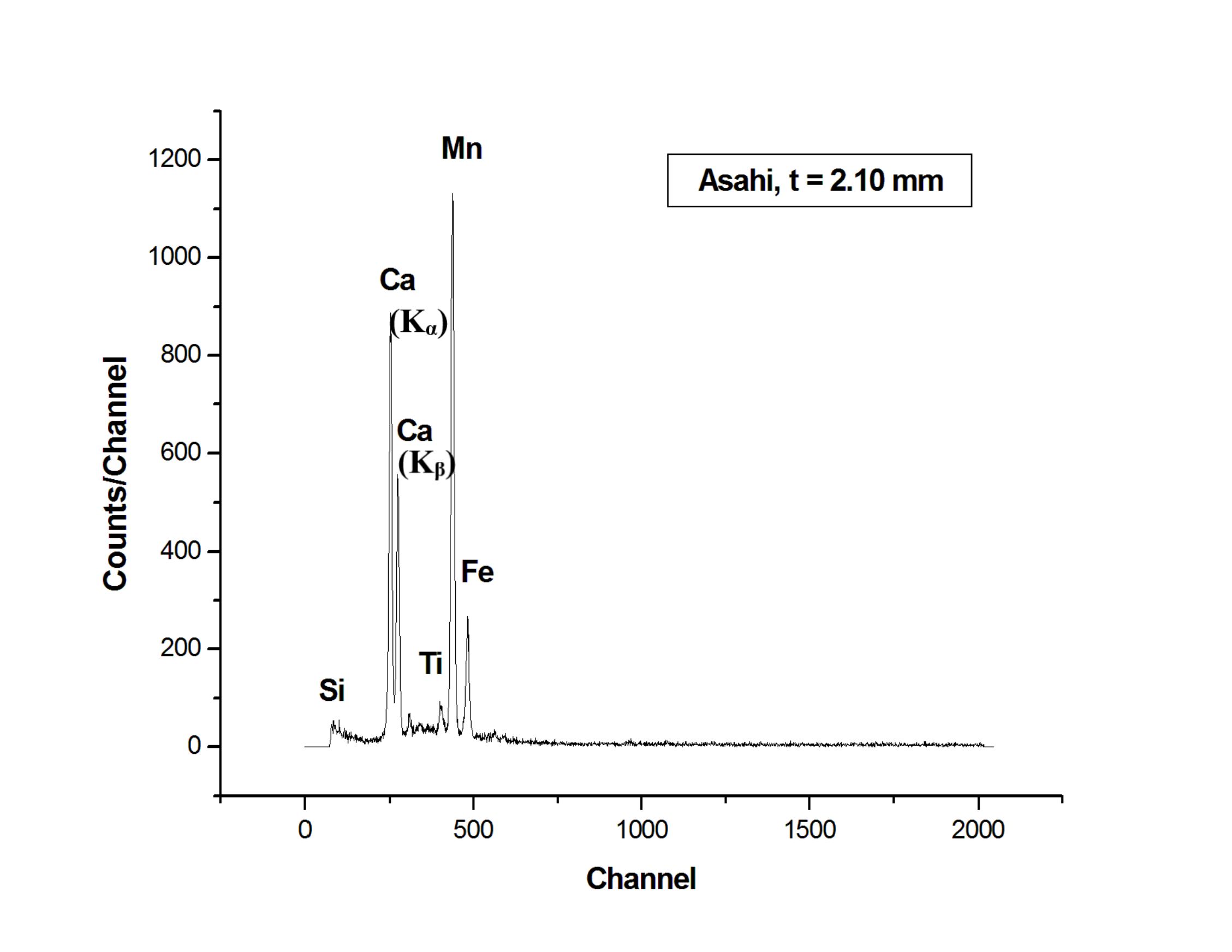}
\includegraphics[height=.3\textheight]{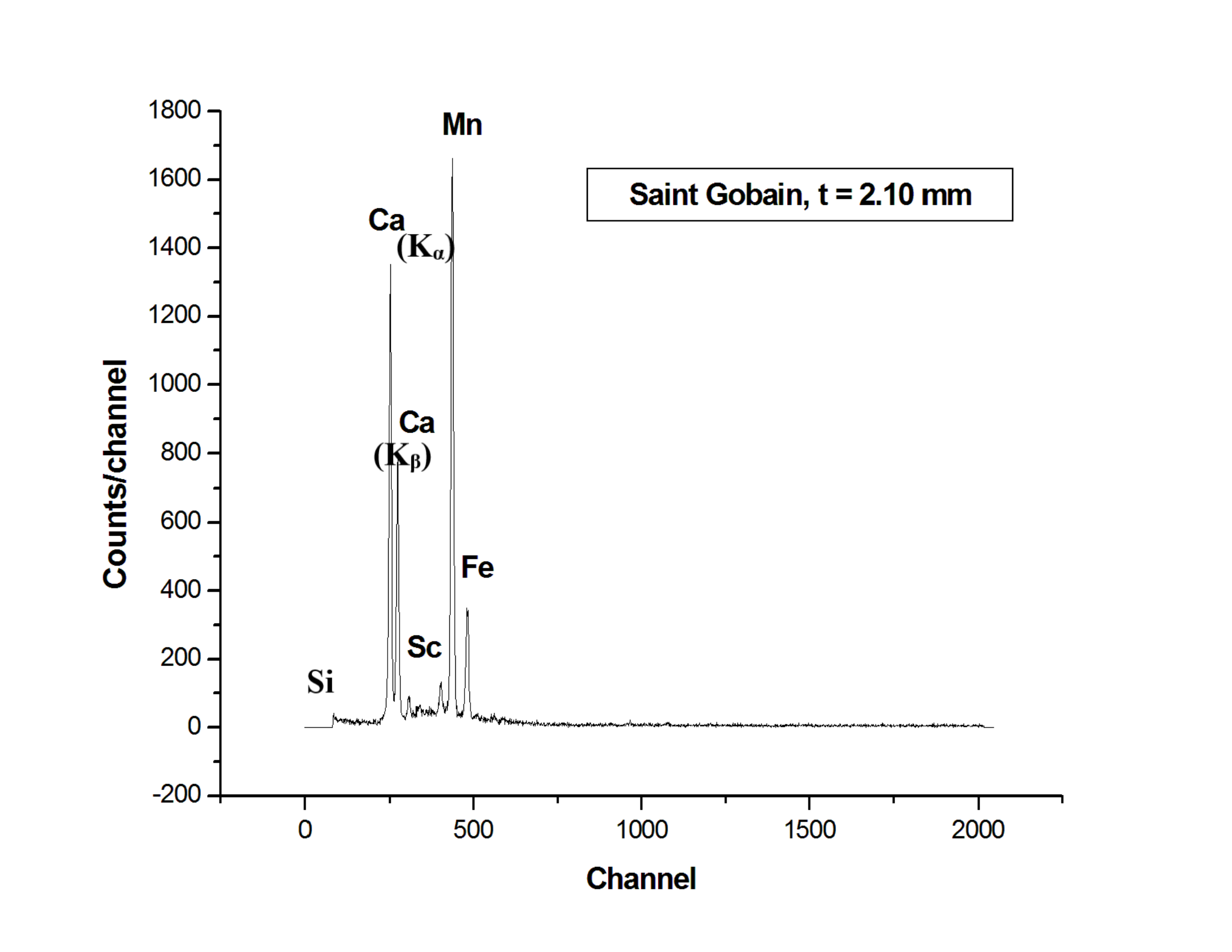}
 \includegraphics[height=.3\textheight]{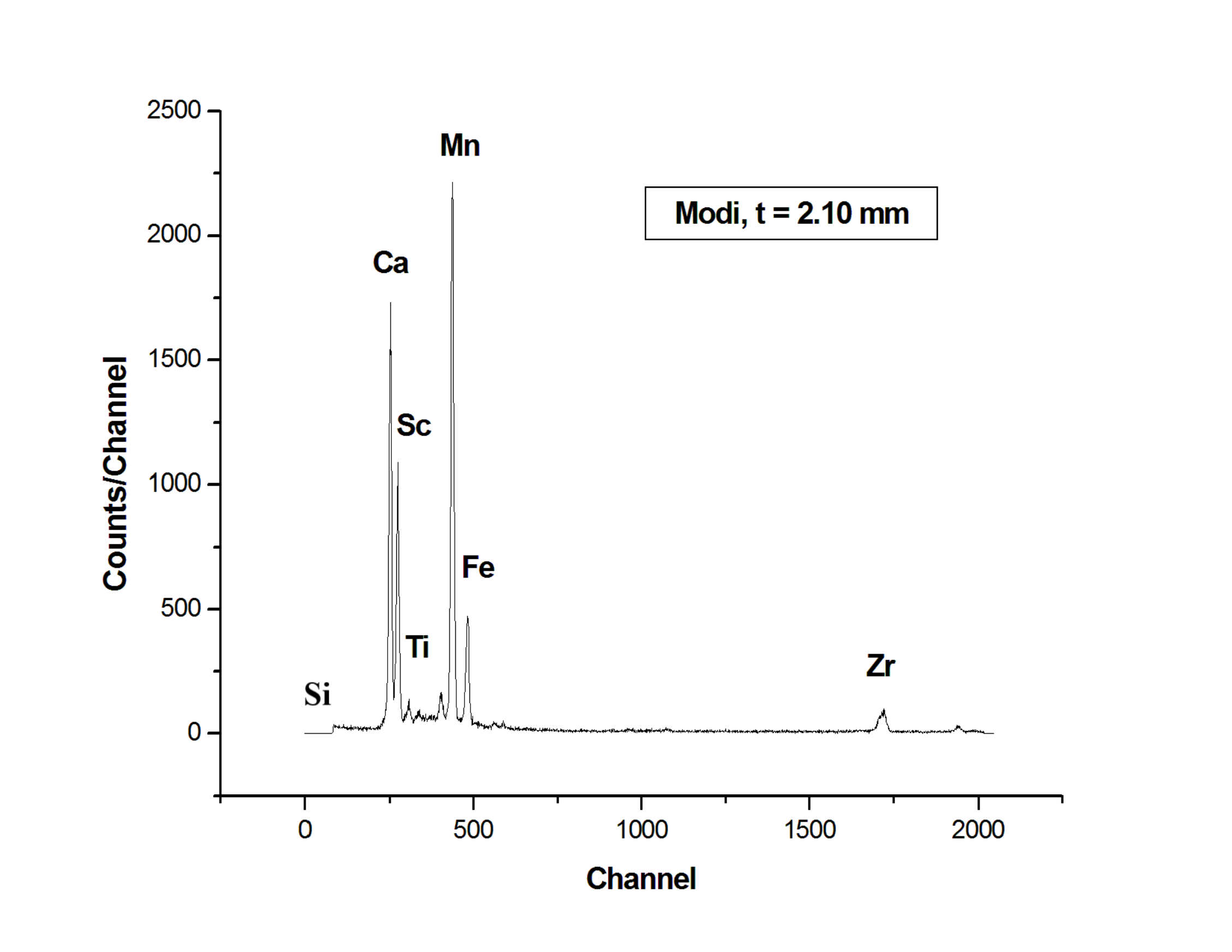}
  \caption{PIXE analysis of A, S and M glass samples respectively.}
\label{fig:PIXE1}
\end{figure}

\noindent
    {\bf {Elemental and Compositional Studies}}: These studies have been performed to check if there was any amount of lead in the glass samples chosen for our study. This is necessary because the presence of a high amount of lead in the glass lowers the melting point and decreases its hardness. That kind of glass has a soft surface and changes the surface quality which is crucial for the electrode study. Wavelength Dispersive X-ray Fluorescence (WD-XRF) spectrum was obtained using Bruker's S8 Tiger WD-XRF spectrometer \cite{cil}. WD-XRF is an analytical method to determine the chemical composition of the materials that can be in solid, liquid, powder, filtered or other form. In this method a high energy electron beam strikes a sample, and photons with a broad continuum of energies are emitted. This is called bremsstrahlung, and gives us a continuous spectrum as shown in figure~\ref{fig:WDXRF11}. Also, interaction between the electron beam and sample causes the ejection of photoelectrons from the inner shells of the atoms. These photoelectrons have a kinetic energy (E-$\phi$), which is the difference between the incident particle energy (E) and the binding energy ($\phi$) of the atomic electron. Thus the ejected electron leaves a hole and another electron from a higher energy shell fills the vacancy. This process is known as {\it fluorescence}. This represents the characteristic spectrum and has been shown as sharp peaks in figure~\ref{fig:WDXRF11}. The quantity of elements were obtained from WD-XRF to get the information of elements and/or ions present in the glass that has been shown in table~\ref{table2}. Another technique called PIXE (Proton Induced X-ray Emission spectroscopy) was used as a supplementary test to WD-XRF. This technique was also used to do the elemental analysis using a cyclotron facility in the Department of Physics, Panjab University Chandigarh. PIXE has an advantage that the cross-section for the production of x-rays is large and the background contribution from bremsstrahlung is low. The energetic protons excite the inner shell electrons in the target atoms and expel them, resulting in the production of x-rays. The energies of these x-rays are unique characteristic of the elements from which they originate and thus the elements were determined. Figures~\ref{fig:WDXRF11} and~\ref{fig:PIXE1} demonstrate the WD-XRF and PIXE analysis of the glass samples respectively. $SiO_{2}$ and $Na_{2}O$ were present in large amounts as they are the main compounds present in the glass. 

Thus, on the basis of optical and surface properties, it was concluded that A and S glass samples are of better quality than the M glass sample. The surface of the M glass sample was observed to be of poor quality on the basis of SEM scans. These properties affect the operation of RPC made of M glass sample. To confirm this we fabricate and characterize the RPCs made from A, S and M glass samples to find the leakage current.

\begin{figure}[htbp!]
\centering
 
\includegraphics[height=12cm, width=14cm, scale=10]{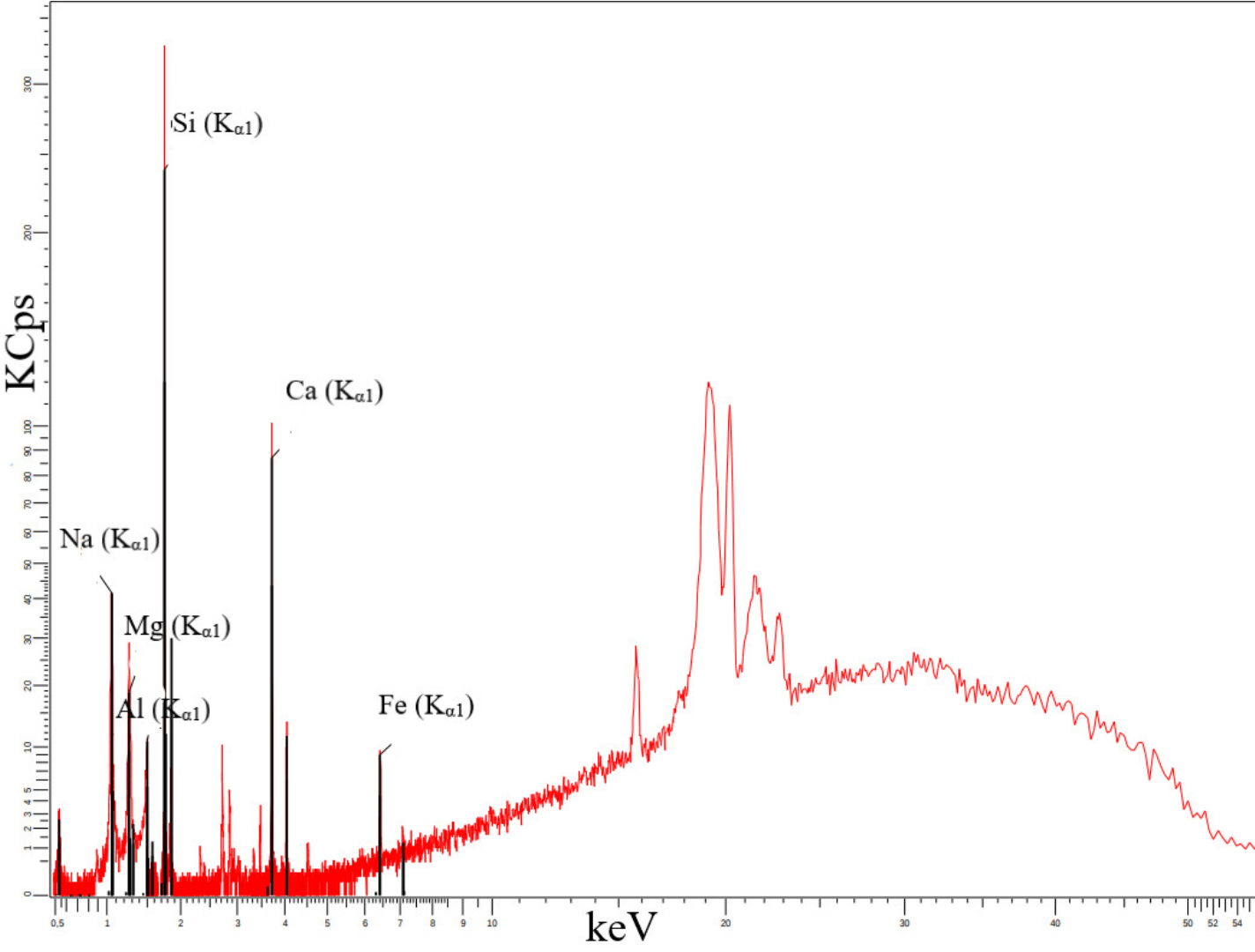}
\includegraphics[height=12cm, width=14cm, scale=10]{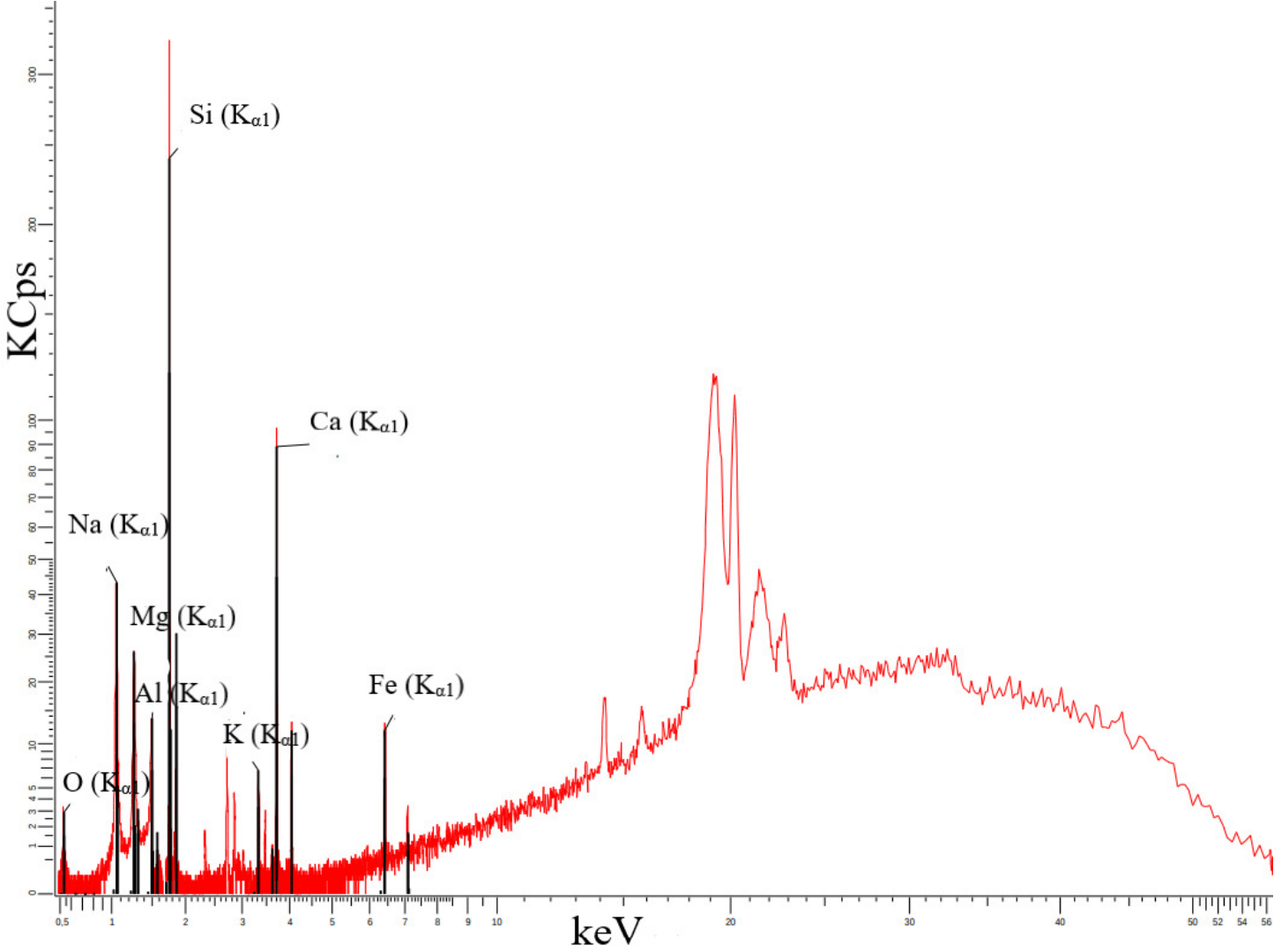}

\end{figure}

\begin{table}[ht]
\centering 
\scalebox{0.8}{
\begin{tabular}{|c c c c|} 
\hline 
Compound & M($\%$) & S($\%$) & A($\%$)\\
\hline                  
$SiO_{2}$ & 72.43 & 73.03 & 71.84 \\ 
$Na_{2}O$ & 13.04 & 13.05 & 14.04 \\
$CaO$ & 9.13 & 8.45 & 8.96 \\
$MgO$ & 3.85 & 3.72 & 4.06  \\
$Al_{2}O_{3}$ & 0.94 & 0.79 & 0.69 \\
$SO_{3}$ & 0.13 & 0.23 & 0.15 \\
$Fe_{2}O_{3}$ & 0.10 & 0.18 & 0.11 \\
$K_{2}O$ & 0.25 & 0.48 & $-$ \\
$TiO_{2}$ & 0.06 & $-$ & 0.09 \\
$ZrO_{2}$ & 99 PPM & 40 PPM & 0.02 \\
$SrO$ & $-$ & 0.01 & $-$ \\
$Sc_{2}O_{3}$ & 0.02 & $-$ & $-$ \\
$Cl$ & $-$ & 0.03 & 0.03 \\
\hline
\end{tabular}
}
\caption{The composition of glass samples from WD-XRF.} 
\label{table2} 
\end{table}

%%%%%%%%%%%%%%%%%%%%%%%%%%%%%%%%%%%%%%%%%%%%%%
%%%%%%%%%%%%%%%%%%%%%%%%%%%%%%%%%%%%%%%%%%%%%%
%\noindent
\section{Fabrication and Characterization of RPC}
\label{fabri}

Before characterizing the glass samples we fabricated RPCs as described in the following section.

\subsection{RPC Assembly Procedure}
\label{assemblyP}
 \noindent
Float glasses of 2.10 mm thickness of A, S and M samples were taken to fabricate RPCs of 30 cm $\times$ 30 cm size with the corners chamfered at $45^{0}$. The glass plates, nozzles, buttons, and side spacers were cleaned with distilled water and propanol. The glass plate was put on a clean white sheet on a table and a drop of epoxy glue was put at four equidistant positions, and one drop was put at the center. Then the button spacers were placed on the top of each glue drop, and pressed lightly to be glued. All the side spacers of 10 mm diameter were then glued at the four corners of the glass and a frame was made that covered all sides of the glass. Nozzles were glued at the corners in such a manner that they were either in a clockwise or anti-clockwise direction. After 12 hours of drying, again a drop of glue was put on each button spacer and then the second glass was placed on top of the button spacers. The glue was applied again on all the sides for the side spacers, nozzles, and glass. The whole set-up was kept for more than 12 hours at room temperature to dry. The glass RPC was then spray painted with the help of a spray gun to apply the Nerolac conduction paint, that had the paint to thinner ratio 1:1. The surface resistance was maintained on the order of 1 M$\Omega$ $\pm$ 200 k$\Omega$. This coating of graphite painting allows uniform application of the high voltage. All the three RPCs were fabricated by the same procedure.

\subsection{V-I Characterization of RPC}
\label{viChar}
\begin{figure}[htbp]
\centering
  \includegraphics[height=.4\textheight]{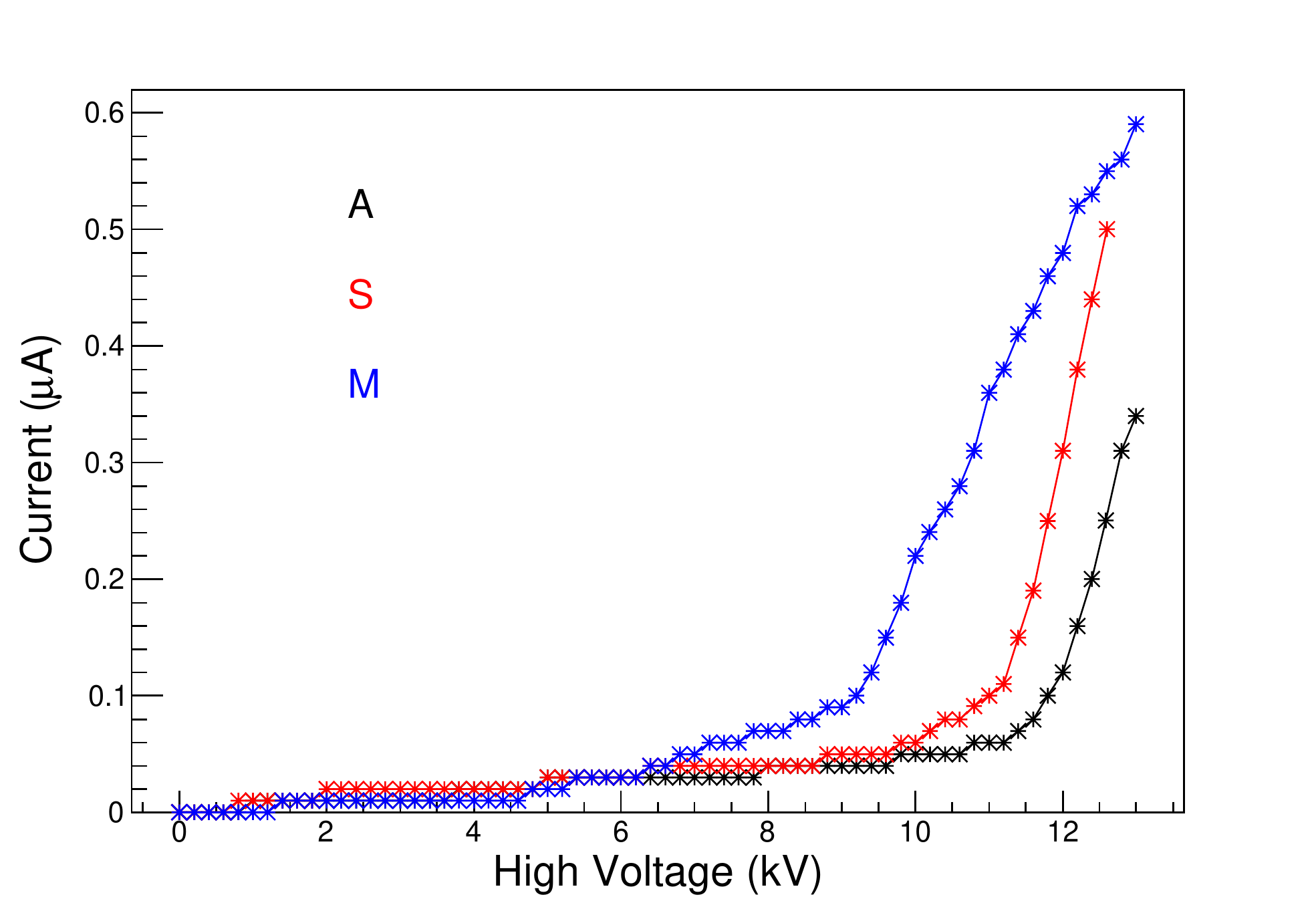}

  \caption{V-I characterization of A, S and M glass samples.}
\label{fig:current}
\end{figure}

 \noindent
 The plot for the V-I characterization was divided into two regimes of voltages: low voltage and high voltage. The choice of the gas mixture depends on factors like low working voltage, high gain, and high rate capability. The three RPCs were operated in the avalanche mode with the mixture of gases \cite{kalmani:2009} in the ratio R134a : Isobutane : $SF_{6}$ :: 95.15 : 4.51 : 0.34. R134a is an eco-friendly substitute of Freon gas \cite{satya}. The mixture of R134a and Isobutane is the quencher and sufficient for the RPC operation. A slight amount of $SF_{6}$ was added to bring stability in the RPC operation. The leakage current for M glass sample was higher than A and S samples as shown in figure~\ref{fig:current}. The high leakage current of M glass sample was in accordance with the surface roughness as discussed in section ~\ref{char_glass}. Thus the M glass sample has the roughest surface among the other two glass samples \cite{kanish}.

\section{Conclusions}
\label{diss_con}

Resistive plate chambers are the main component of the ICAL and many particle physics detectors. In ICAL detector at INO there would be $\sim$ 30000 glass RPCs, so their R \& D is necessary. Glass is an important component and can affect the dynamics of the whole detector that further affects the physics goals to be obtained with the help of these detectors. So detailed studies should be done before making a final decision. The glass samples named A, S and M were procured from a local market. Their physical, electrical, and optical properties, surface characteristics, elemental composition, and V-I characterization of RPCs made from the chosen glass samples were studied. The various techniques were used to study the different properties which helped to distinguish the different glass samples. The surface quality is the main factor in deciding the best among them. The rough surface of these glass electrodes cause the electric field of the RPC to vary significantly. If the surface is rough it will form a discharge and hence affect the operation and stabilization of the RPCs. SEM scans have enabled to view the roughness of the glass sample which further motivated to check the operation of RPCs. The WD-XRF and PIXE tests were done as an addition to check the presence or absence of lead quantity in the samples. The RPCs were fabricated with the same glass materials and tested for the V-I characterization to confirm the facts about the glass surface roughness. A comparative scale to choose the best glass sample as an electrode for RPC was found. On the basis of their optical, surface properties, and V-I characterization it was concluded that A and S glass samples are better electrodes than M glass sample.

\paragraph{Acknowledgements}

The authors thank CIL department of Panjab University (PU) and Cyclotron facility in Physics Department of Panjab University for characterization techniques. We thank the technical staff of the EHEP group in the Physics department. R. Kanishka acknowledges UGC/DAE/DST (Govt. of India) for funding.


\begin{thebibliography}{99}

\bibitem{cardarelli} R. Santonico, R. Cardarelli, { \it Development of resistive plate counters}, \href{https://doi.org/10.1016/0029-554X(81)90363-3}{{\it Nucl. Instrum. Meth. A} {\bf 187} (1981) 377--380}.

  \bibitem{Santonico} R. Cardarelli et al., {\it Progress in resistive plate counters}, \href{https://doi.org/10.1016/0168-9002(88)91011-X}{{\it Nucl. Instrum. Meth. A} {\bf 263} (1988) 20--25}.
 
\bibitem{satya} B. Satyanarayana, {\it Design and Characterisation Studies of Resistive Plate Chambers, Ph.D thesis}, Department of Physics IIT Bombay, PHY-PHD-10-701 (2009).
  
\bibitem{belle} Abe T. et al., Belle-II Collaboration, {\it Belle II Technical Design Report}, KEK-REPORT-2010-1 (2010). \href{http://arxiv.org/abs/1011.0352}{[arxiv:1011.0352]}.

\bibitem{cms} CMS Collaboration, {\it The Compact Muon Solenoid: Technical proposal}, CERN-LHCC-94-38;LHCC-P-1 (1994).
  
\bibitem{ino} S. Atthar et al., (INO Collaboration), INO Project Report, INO/2006/01, June 2006, \href{http://www.ino.tifr.res.in/ino/OpenReports/INOReport.pdf}{http://www.ino.tifr.res.in/ino/OpenReports/INOReport.pdf}.
  
\bibitem{fonte} P. Fonte, {\it High-resolution timing of MIPs with RPCs - a model}, \href{https://doi.org/10.1016/S0168-9002(00)00953-0}{{\it Nucl. Instrum. Meth. A} {\bf 456} (2000) 6}.
\bibitem{carda1} R. Cardarelli, R. Santonico et. al., {\it Avalanche and streamer mode operation of resistive plate chambers}, \href{https://doi.org/10.1016/S0168-9002(96)00811-X}{{\it Nucl. Instrum. Meth. A} {\bf 382} (1996) 470--474}.

\bibitem{muon1} Seth H. Neddermeyer and Carl D. Anderson, {\it Note on the Nature of Cosmic-Ray Particles}, {\it Phys. Rev.} {\bf 51(10)} (1937) 884--886.
  \href{https://link.aps.org/doi/10.1103/PhysRev.51.884}{https://link.aps.org/doi/10.1103/PhysRev.51.884}.

\bibitem{Gasik} P. Gasik et al., {\it Charge density as a driving factor of discharge formation in GEM-based detectors}, {\it Nucl. Instrum. Meth. A} {\bf 870} (2017) 116--122. \href{https://arxiv.org/abs/1704.01329}{[arxiv:1704.01329]}.

\bibitem{Meghna} K. K. Meghna, et al., {\it Measurement of electrical properties of electrode materials for the bakelite Resistive Plate Chambers}, {\it JINST} {\bf 7} P10003 (2012).
  
\bibitem{kanish} R. Kanishka et al., {\it Optimisation and Characterisation of Glass RPC for India-based Neutrino Observatory Detectors}. \href{https://arxiv.org/abs/1605.09361}{[arxiv:1605.09361]}.

\bibitem{sgobain} https://uk.saint-gobain-building-glass.com/en-gb/architects/physical-properties
  
 \bibitem{cil} https://onlinesaif.puchd.ac.in/
 
 \bibitem{vaccum} Goldstein J.I. et al., {\it Coating and Conductivity Techniques for SEM and Microanalysis. In: Scanning Electron Microscopy and X-Ray Microanalysis}, {Springer, Boston, MA} (1992) 671--740. \href{https://doi.org/10.1007/978-1-4613-0491-3\_13}{https://doi.org/10.1007/978-1-4613-0491-3\_13}.
   \bibitem{kalmani:2009} S. D. Kalmani et al., {\it On-line gas mixing and multi-channel distribution system}, {\it Nucl. Instrum. Meth. A}, {\bf 602} (2009) 845--849. \href{https://doi.org/10.1016/j.nima.2008.12.153}{https://doi.org/10.1016/j.nima.2008.12.153}.
\end{thebibliography}
\end{document}